\documentclass[twoside]{articlek}
\usepackage{amssymb}
\usepackage{cite}

\renewcommand{\refname}{REFERENCES}
\def\be{\begin{equation}}
\def\ee{\end{equation}}

\def\BibTeX{{\rm B\kern-.05em{\sc i\kern-.025em b}\kern-.08em
            T\kern-.1667em\lower.7ex\hbox{E}\kern-.125emX}}

\usepackage{graphicx}
\begin{document}
\sloppy
\twocolumn[{
{\large\bf MAGNETIC ANALOGUE OF LIQUID-GAS PHASE TRANSITION OF WATER: \\
CASE STUDY OF A SPIN-1/2 ISING-HEISENBERG MODEL ON A DIAMOND-DECORATED SQUARE LATTICE}\\
{\small J.\ Stre\v{c}ka, jozef.strecka@upjs.sk, K.\ Karl'ov\'a, katarina.karlova@upjs.sk, Institute of Physics, Faculty of Science, P.~J.\ \v{S}af\'arik University, Park Angelinum 9, 04001 Ko\v{s}ice, Slovakia, T.\ Verkholyak, werch@icmp.lviv.ua, Institute for Condensed Matter Physics, NASU, Svientsitskii Street 1, 790 11, L'viv, Ukraine, N.\ Caci, caci@physik.rwth-aachen.de, S.\ Wessel,
wessel@physik.rwth-aachen.de,
Institute for Theoretical Solid State Physics, JARA FIT, and JARA CSD, RWTH Aachen University, 52056 Aachen, Germany and A.\
Honecker, andreas.honecker@cyu.fr, Laboratoire de Physique Th\'eorique et Mod\'elisation, CNRS UMR 8089, CY Cergy Paris Universit\'e, Cergy-Pontoise, France}\\


}]
\section{INTRODUCTION}

Phase transitions of diverse physical systems have captured significant
attention due to the occurrence of abrupt discontinuities or divergences in several physical quantities, which consequently preclude the proper definition at the relevant phase transitions \cite{dom72}. One of the most extensively studied physical substances is water, whose phase diagram encompasses solid, liquid, and gaseous phases separated from each other by
lines of discontinuous phase transitions. Among these phase-transition lines, the most intriguing is the line of liquid-gas phase transitions, which starts at a triple coexistence point of all three phases and ends at the critical point where a discontinuous phase transition changes to a continuous one. A similar phase-transition line has been recently found in the pressure-temperature phase diagram of the quantum magnetic material SrCu$_2$(BO$_3$)$_2$, which provides an experimental realization of the spin-1/2 Heisenberg model on the Shastry-Sutherland lattice \cite{lar21}.  

The line of discontinuous phase transitions terminating at an Ising critical point is not unique to the Shastry-Sutherland model, but it may be encountered in
other frustrated two-dimensional quantum spin systems such as the spin-1/2 Heisenberg model on a fully frustrated bilayer \cite{sta18,web22,fan23}, trilayer \cite{hon22} and diamond-decorated square lattice \cite{cac23}. To capture thermal phase transitions of the spin-1/2 Heisenberg model on frustrated two-dimensional lattices one has to resort to state-of-the-art numerical calculations as for instance quantum Monte Carlo simulations in a dimer or trimer basis in order avoid the notorious sign problem \cite{gub16}. Interestingly, it has been recently verified that the simpler spin-1/2 Ising-Heisenberg model on the diamond-decorated square lattice already captures the essential features of thermal phase transitions of the spin-1/2 Heisenberg model on the diamond-decorated square lattice \cite{cac23,str23}.

The spin-1/2 Ising-Heisenberg diamond-decorated square lattice in a magnetic field can be rigorously mapped to an effective zero-field spin-1/2 Ising square lattice in a particular parameter subspace corresponding to the thermal phase transitions \cite{str23}. From this perspective, the spin-1/2 Ising-Heisenberg model on the diamond-decorated square lattice represents a valuable example in the realm of exactly solved lattice-statistical models \cite{bax82}, which allows us to rigorously examine the thermal phase transitions in the presence of an external magnetic field quite similarly as for Fisher's superexchange antiferromagnet \cite{fis60} and its different variants \cite{hat68,mas73,luw05,can06,gal16}.

In the present paper we provide a more comprehensive understanding of the thermal phase transitions of the spin-1/2 Ising-Heisenberg diamond-decorated square lattice. Except temperature and magnetic-field variations of the local and total magnetization, we will bring insight into the respective changes of the magnetic susceptibility and specific heat at the discontinuous as well as continuous thermal phase transitions. 

The paper is organized as follows. In the following section we will briefly recall the definition of the studied quantum spin model and review the mapping to the effective spin-1/2 Ising model on a square lattice. The subsequent section is devoted to a detailed description of basic magnetic and thermodynamic quantities (magnetization, susceptibility, and specific heat) in close vicinity of the thermal phase transitions. Finally, the most important findings are summarized in the concluding part.

\section{ISING-HEISENBERG DIAMOND-DE\-CO\-RA\-TED SQUARE LATTICE}

\begin{figure}[t!]
\begin{center}
\includegraphics[width=0.3\textwidth]{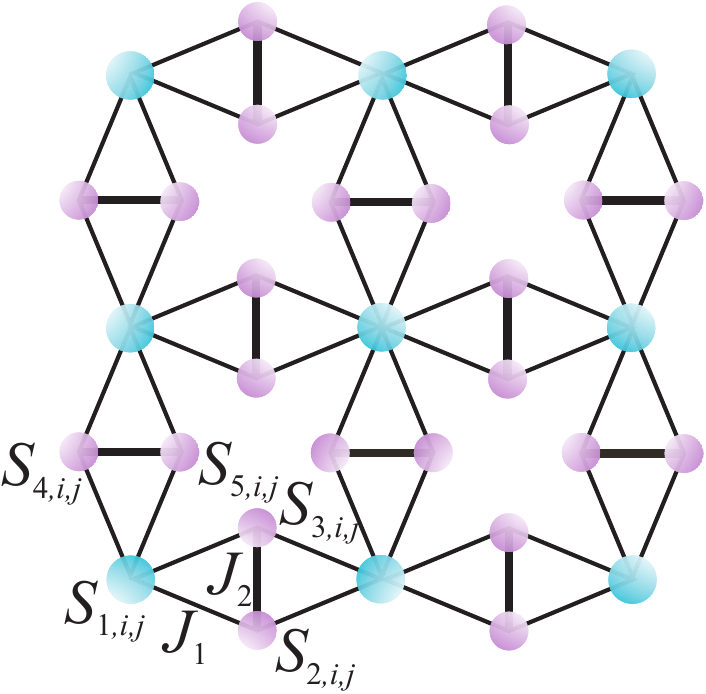}
\end{center}
\vspace{-0.5cm}
\caption{Schematic illustration of the spin-1/2 Ising-Heisenberg model on a diamond-decorated square lattice. Large (light blue) circles determine nodal lattice sites occupied by the Ising spins $S_{1,i,j}$, while small (violet) circles determine decorating lattice sites occupied by the Heisenberg spins $S_{k,i,j}$ ($k=2-5$).}
\label{lattice}       
\end{figure}

At first, let us recall the definition of the spin-1/2 Ising-Heisenberg model on the diamond-decorated square lattice, which is schematically illustrated in Fig.~\ref{lattice} and mathematically given by the Hamiltonian
\begin{eqnarray}
\hat{H} \!\!\!&=&\!\!\! J_1 \sum_{i=1}^{L} \sum_{j=1}^{L} 
\Bigl[ \left(\hat{{S}}_{1,i,j}^z + \hat{{S}}_{1,i+1,j}^z \right) \!
\left(\hat{{S}}^z_{2,i,j} + \hat{{S}}^z_{3,i,j} \right) \nonumber \\ 
&& \qquad \quad + \left(\hat{{S}}_{1,i,j}^z + \hat{{S}}_{1,i,j+1}^z \right) \!
\left(\hat{{S}}^z_{4,i,j}+\hat{{S}}^z_{5,i,j}\right) \Bigr] \nonumber \\
&&+ J_2 \sum_{i=1}^{L} \sum_{j=1}^{L} 
\left(\hat{\bf {S}}_{2,i,j}\cdot\hat{\bf {S}}_{3,i,j} 
+ \hat{\bf {S}}_{4,i,j}\cdot\hat{\bf {S}}_{5,i,j}\right) \nonumber \\  
&&- h \sum_{k=1}^{5} \sum_{i=1}^{L} \sum_{j=1}^{L} \hat{S}_{k,i,j}^z. 
\label{eq:ham}
\end{eqnarray}
The first term accounts for an anisotropic (Ising-type) exchange interaction $J_1$ between nearest-neighbor Ising and Heisenberg spins schematically shown in Fig.~\ref{lattice} as large and small circles, respectively, while the second term stands for the isotropic exchange interaction between nearest-neighbor Heisenberg spins. The last term takes	into account the Zeeman energy of the Heisenberg and Ising spins in an external magnetic field $h$ and $L$ denotes the linear size of the considered two-dimensional lattice. 

It has been verified in our previous paper \cite{str23} that the partition function of the spin-1/2 Ising-Heisenberg model on the diamond-decorated square lattice given by the Hamiltonian (\ref{eq:ham}) can be related via the generalized decoration-iteration transformation \cite{fis59,roj09,str10} to the partition function of the effective spin-1/2 Ising model on a square lattice:
\begin{eqnarray}
{Z} (\beta, J_1, J_2, h) =  A^{2N} {Z}_{\rm eff} (\beta, J_{\rm eff}, h_{\rm eff}), 
\label{zm}
\end{eqnarray}  
which is defined through the effective Hamiltonian involving temperature-dependent nearest-neighbor interactions $J_{\rm eff}$ and magnetic field $h_{\rm eff}$:
\begin{eqnarray}
{H}_{\rm eff} = \!\!\!&-&\!\!\! J_{\rm eff} \sum_{i=1}^L \sum_{j=1}^L (S_{1,i,j}^z S_{1,i+1,j}^z + S_{1,i,j}^z S_{1,i,j+1}^z) \nonumber \\
                \!\!\!&-&\!\!\! h_{\rm eff} \sum_{i=1}^L \sum_{j=1}^L S_{1,i,j}^z. 
\label{hef}										
\end{eqnarray} 
An explicit form of the mapping parameters $A$, $J_{\rm eff}$, and $h_{\rm eff}$ is given by Eqs.~(7)--(10) in Ref.~\cite{str23}. It follows from the exact mapping between the partition functions (\ref{zm}) that the spin-1/2 Ising-Heisenberg model on the diamond-decorated square lattice becomes exactly soluble in the particular parameter subspace with zero effective field $h_{\rm eff} = 0$ due to the famous exact solution by Onsager \cite{ons44}, while in the parameter space with nonzero effective field $h_{\rm eff} \neq 0$ one may obtain precise numerical results by exploiting classical Monte Carlo simulations.  

Except for a trivial fully saturated paramagnetic phase, the spin-1/2 Ising-Heisenberg model on the diamond-decorated square lattice displays two further ground states, which can be classified as the classical ferrimagnetic (FRI) phase:
\begin{eqnarray}
|{\rm FRI} \rangle = \prod_{i,j=1}^L \! |\!\!\downarrow_{1,i,j}\rangle \!\otimes\! |\!\!\uparrow_{2,i,j}\uparrow_{3,i,j}\rangle \!\otimes\! |\!\!\uparrow_{4,i,j}\uparrow_{5,i,j}\rangle
\label{fri}
\end{eqnarray}
and the quantum monomer-dimer (MD) phase:
\begin{eqnarray}
|{\rm MD} \rangle = \prod_{i,j=1}^L \!\!\!\!\!\!\!&&\!\!\!\!\!\!\! |\!\!\uparrow_{1,i,j}\rangle \otimes 
 \!\frac{1}{\sqrt{2}}(|\!\!\uparrow_{2,i,j}\downarrow_{3,i,j}\rangle - |\!\!\downarrow_{2,i,j}\uparrow_{3,i,j}\rangle) \nonumber \\
\!\!\!\!&\otimes&\!\!\!\! \!\frac{1}{\sqrt{2}}(|\!\!\uparrow_{4,i,j}\downarrow_{5,i,j}\rangle - |\!\!\downarrow_{4,i,j}\uparrow_{5,i,j}\rangle). \label{md}
\end{eqnarray} 
At zero temperature the FRI and MD ground states coexist 
along the phase boundary given by:
\begin{eqnarray}
\label{h_boundary}
h_\mathrm{MD-FRI} = 2(J_2 - J_1), \qquad J_2 \in [ J_1, 3 J_1].
\label{Eq:hMDFRI}
\end{eqnarray}
The ground-state phase boundary (\ref{Eq:hMDFRI}) represents a lower boundary for the wall of discontinuous (first-order) phase transitions between the FRI and MD phases, which are bounded from above by a line of continuous (second-order) phase transitions from the universality class of the two-dimensional Ising model \cite{ons44} as dictated by the constraint $\beta_{\rm c} J_{\rm eff} = 2 \ln (1 + \sqrt{2})$ and $h_{\rm eff} = 0$. 

\section{THERMAL PHASE TRANSITIONS BETWEEN FERRIMAGNETIC AND MONO\-MER-DIMER PHASES}

In this section our particular attention will be focused on typical signatures of discontinuous and continuous thermal phase transitions between the FRI and MD phases. To this end, we will limit our further analysis to the spin-1/2 Ising-Heisenberg model on the diamond-decorated square lattice with the specific value of the interaction ratio $J_2/J_1 = 1.5$, which captures all typical features of both types of thermal phase transitions emergent in the close vicinity of the coexistence point between the FRI and MD ground states ($h_\mathrm{MD-FRI}/J_1 = 1$ for $J_2/J_1 = 1.5$).

\begin{figure}[t!]
\includegraphics[width=\columnwidth]{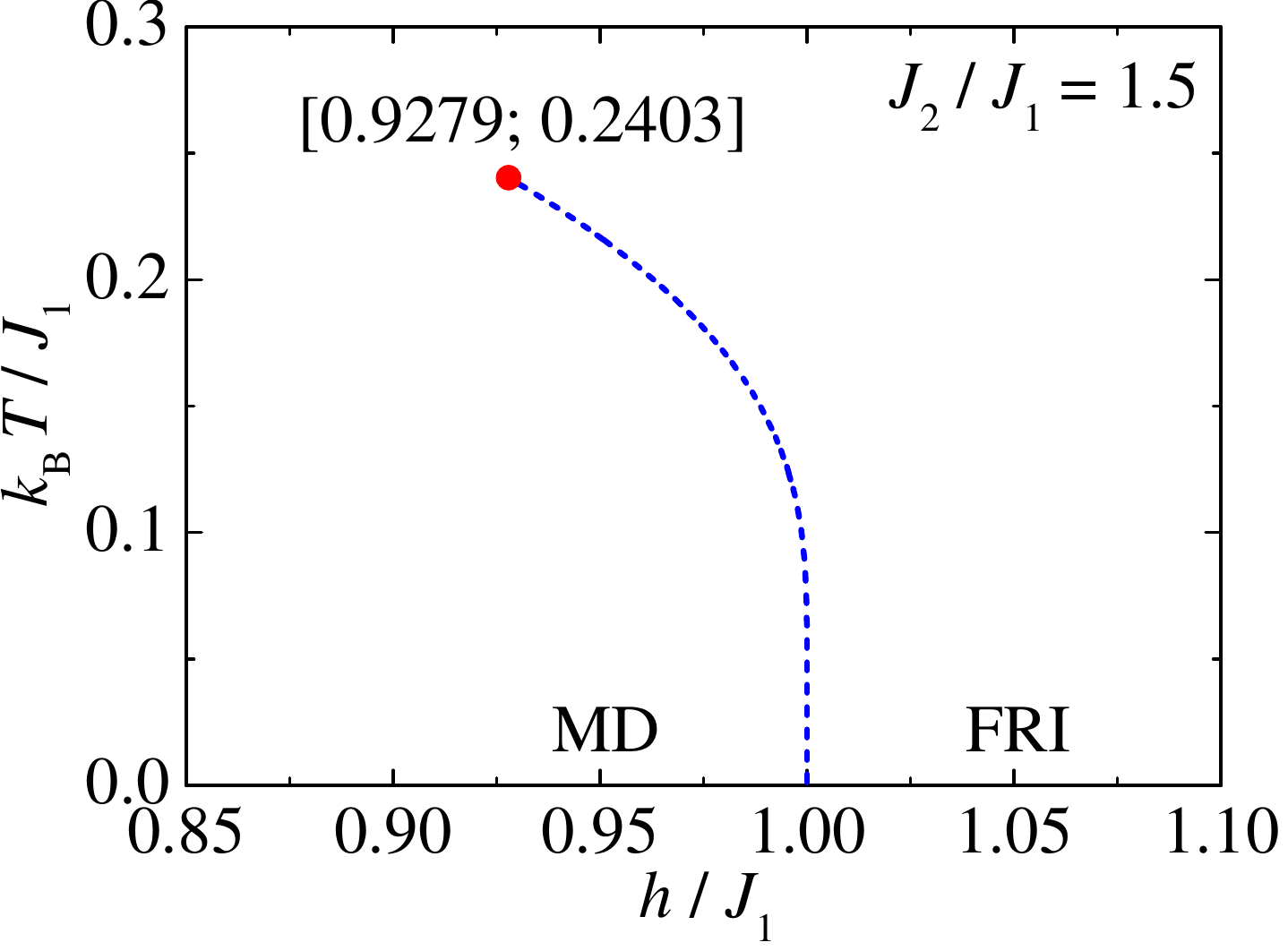}
\caption{Finite-temperature phase diagram 
in the magnetic field versus temperature plane for the fixed value of the interaction ratio $J_2/J_1 =1.5$. The blue broken curve displays the line of discontinuous thermal phase transitions between the quantum monomer-dimer (MD) phase and the classical ferrimagnetic (FRI) phase, which terminates at the Ising critical point (red circle) with the coordinates $h/J_1 = 0.9279$ and  $k_{\rm B} T/J_1 = 0.2403$.}
\label{pd}       
\end{figure}

Figure \ref{pd} reports the finite-temperature phase diagram of the spin-1/2 Ising-Heisenberg model on the diamond-like decorated lattice in the magnetic field versus temperature plane for this ratio. It can be seen from this figure that the line of discontinuous thermal phase transitions between the FRI and MD phases (blue broken line) actually starts from the relevant coexistence point $h/J_1 = 1$ before it bends towards lower magnetic fields as temperature rises. This line of discontinuous thermal phase transitions finally terminates at the Ising critical point at $h/J_1 = 0.9279$ and $k_{\rm B} T/J_1 = 0.2403$ for $J_2/J_1 =1.5$, which can be ascribed to a continuous thermal phase transition between the FRI and MD phases (red circle).  

\begin{figure}[t!]
\includegraphics[width=\columnwidth]{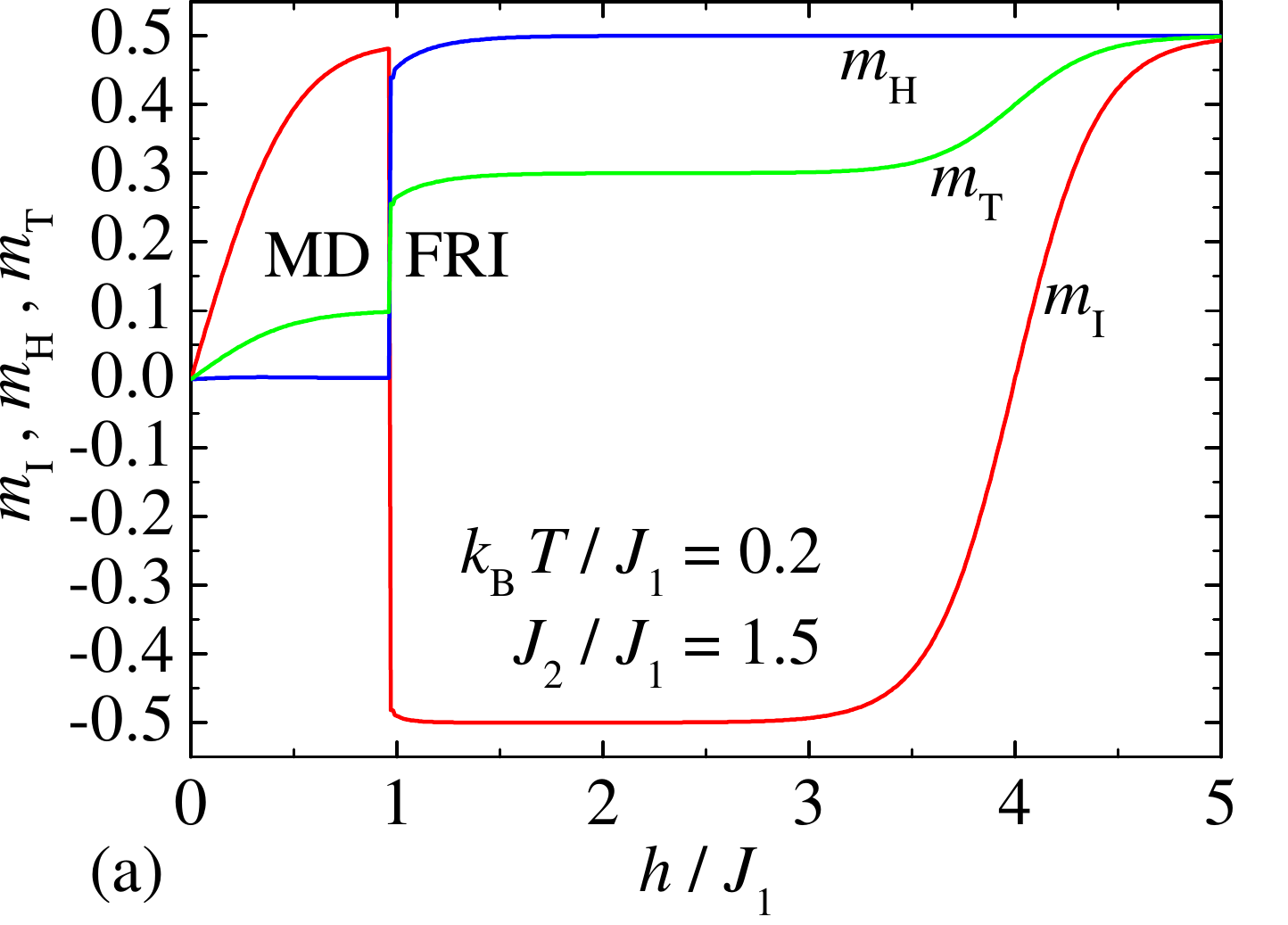}\\[1.5mm]
\includegraphics[width=\columnwidth]{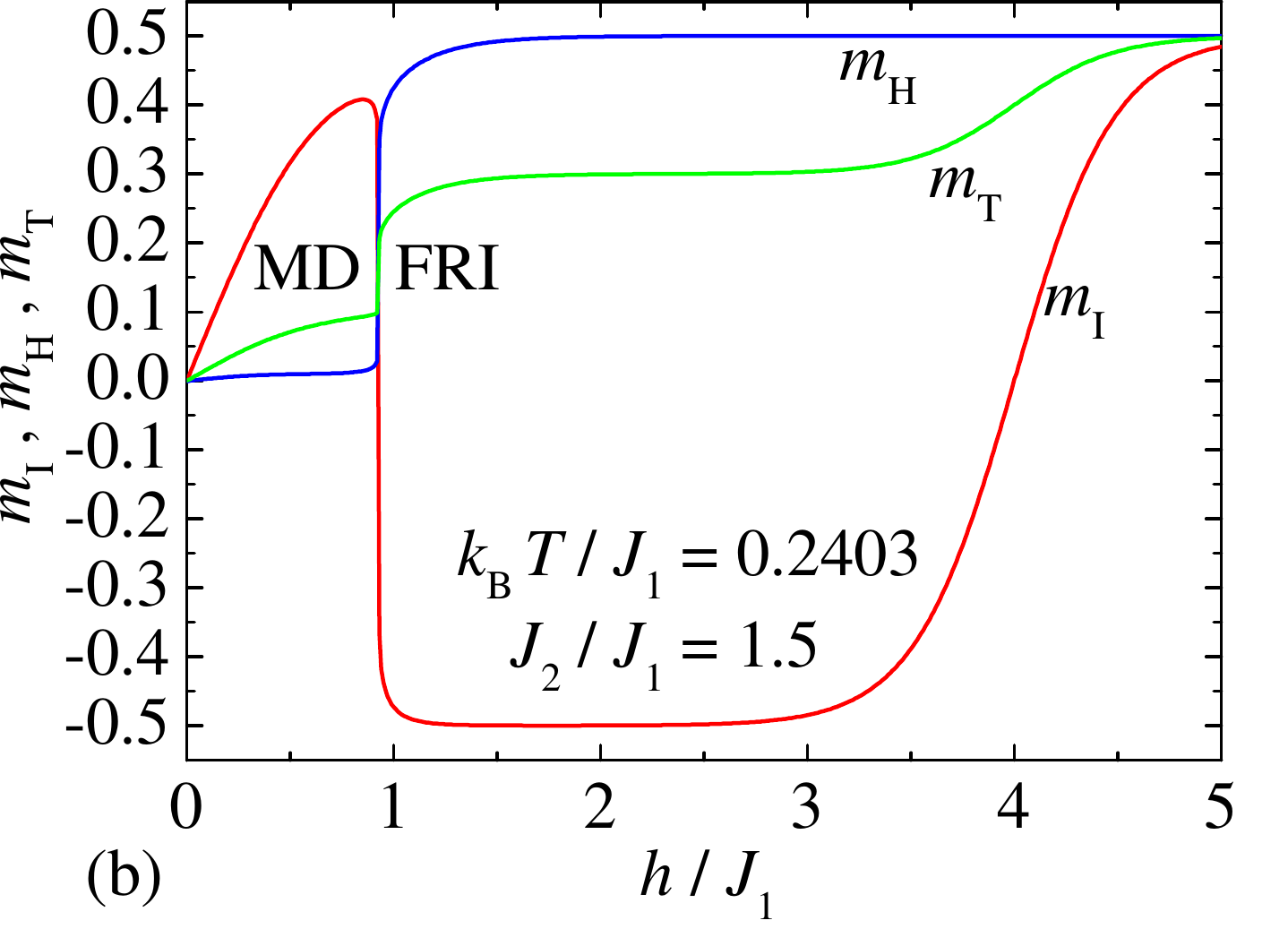}\\[1.5mm]
\includegraphics[width=\columnwidth]{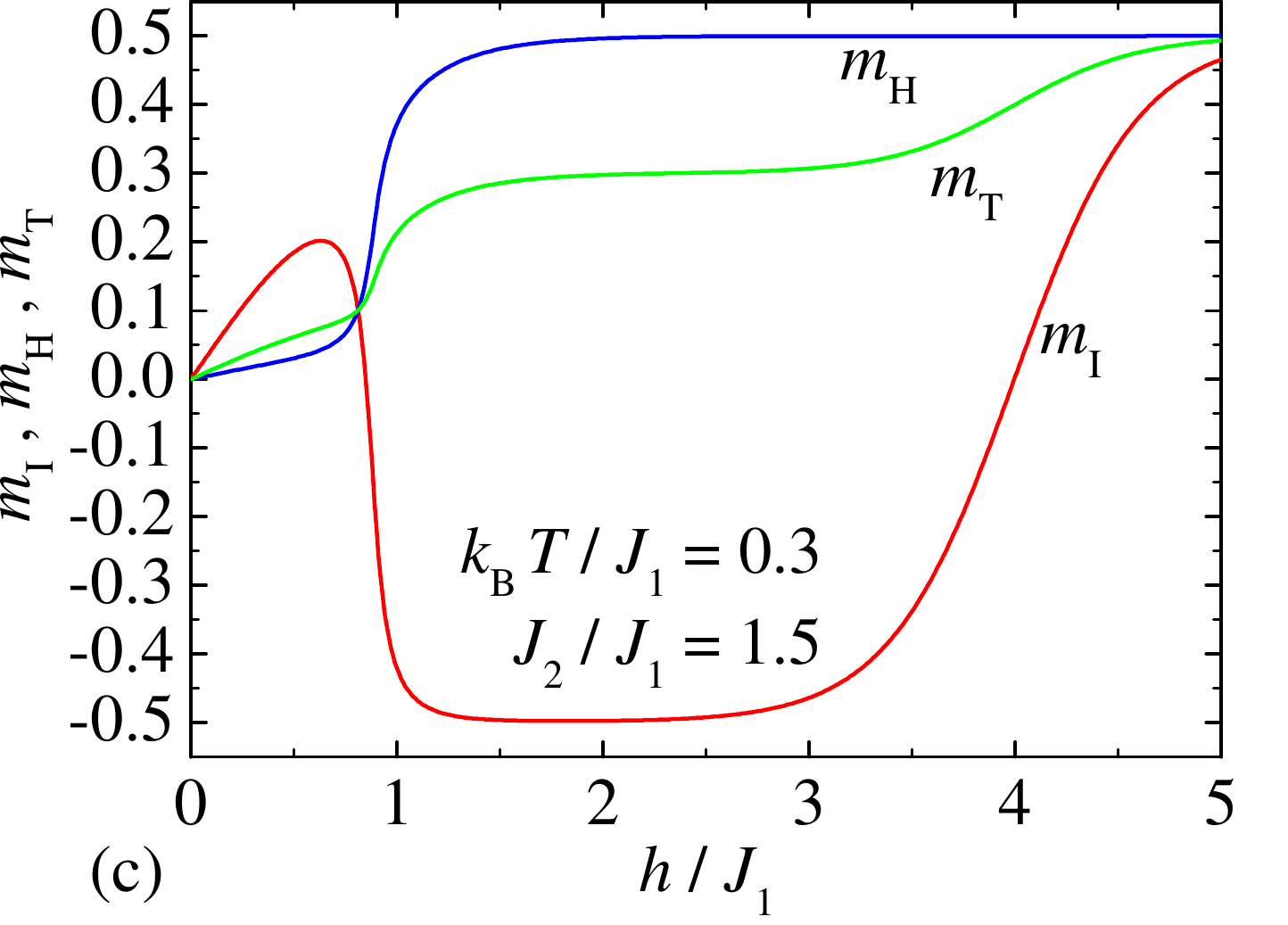}
\caption{A few typical magnetic-field dependencies of the local and total magnetization 
for the fixed value of the interaction parameter $J_2/J_1 =1.5$ and three selected values of temperature:
(a) $k_{\rm B}T/J_1 = 0.2$;
(b) $k_{\rm B}T/J_1 = 0.2403$;
(c) $k_{\rm B}T/J_1 = 0.3$.}
\label{mh}       
\end{figure}

\begin{figure}[t!]
\includegraphics[width=\columnwidth]{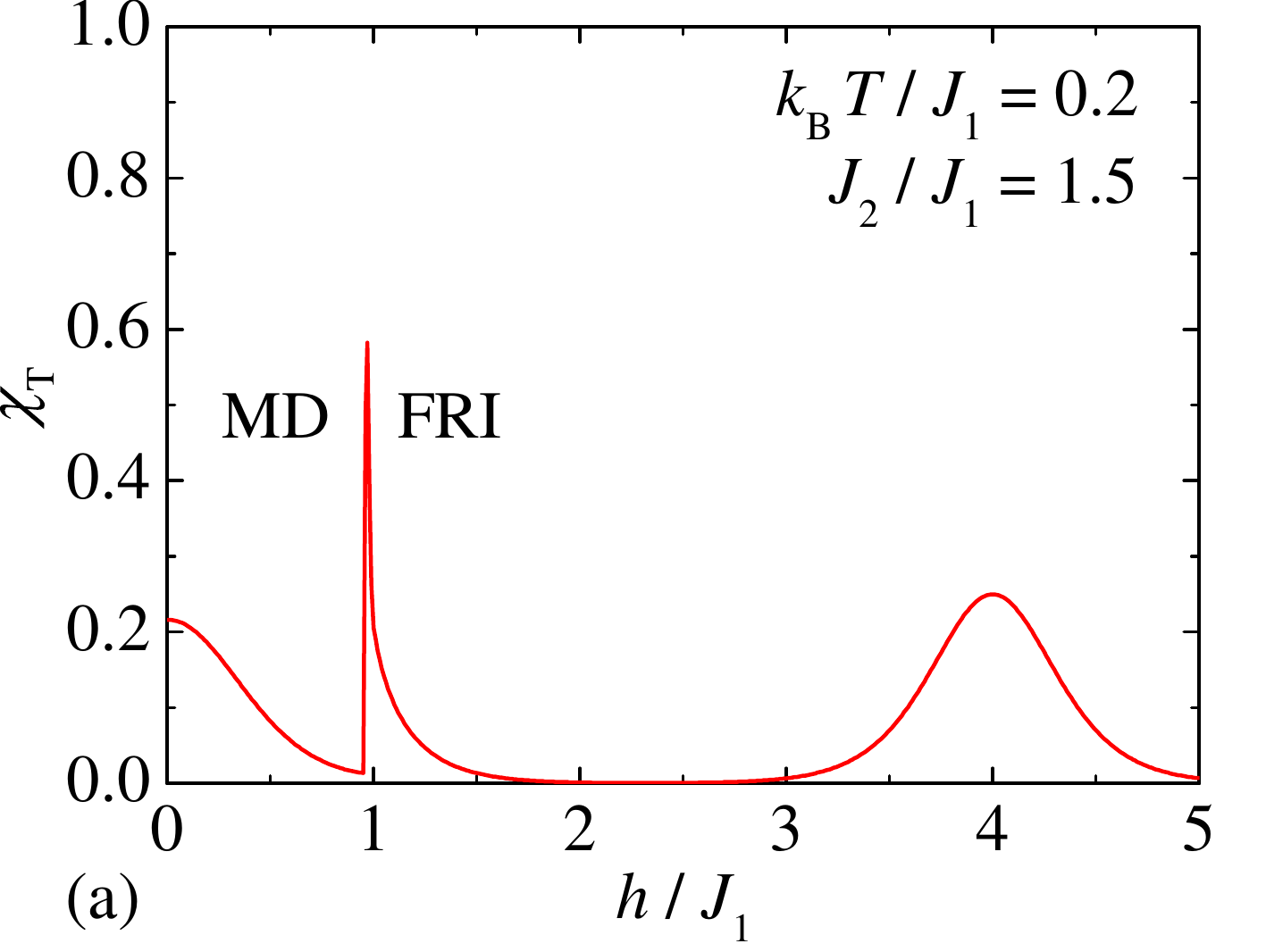}\\[1.5mm]
\includegraphics[width=\columnwidth]{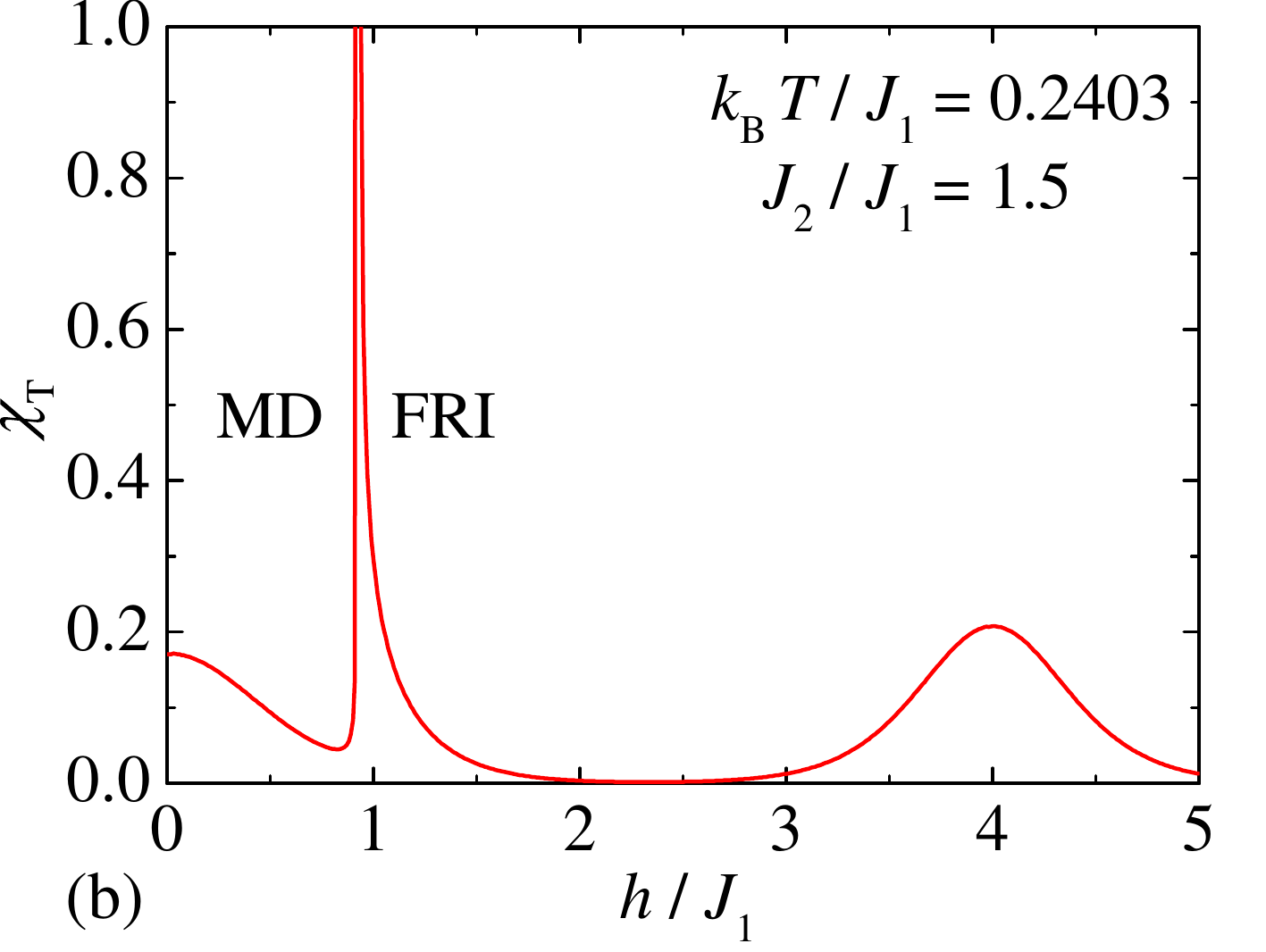}\\[1.5mm]
\includegraphics[width=\columnwidth]{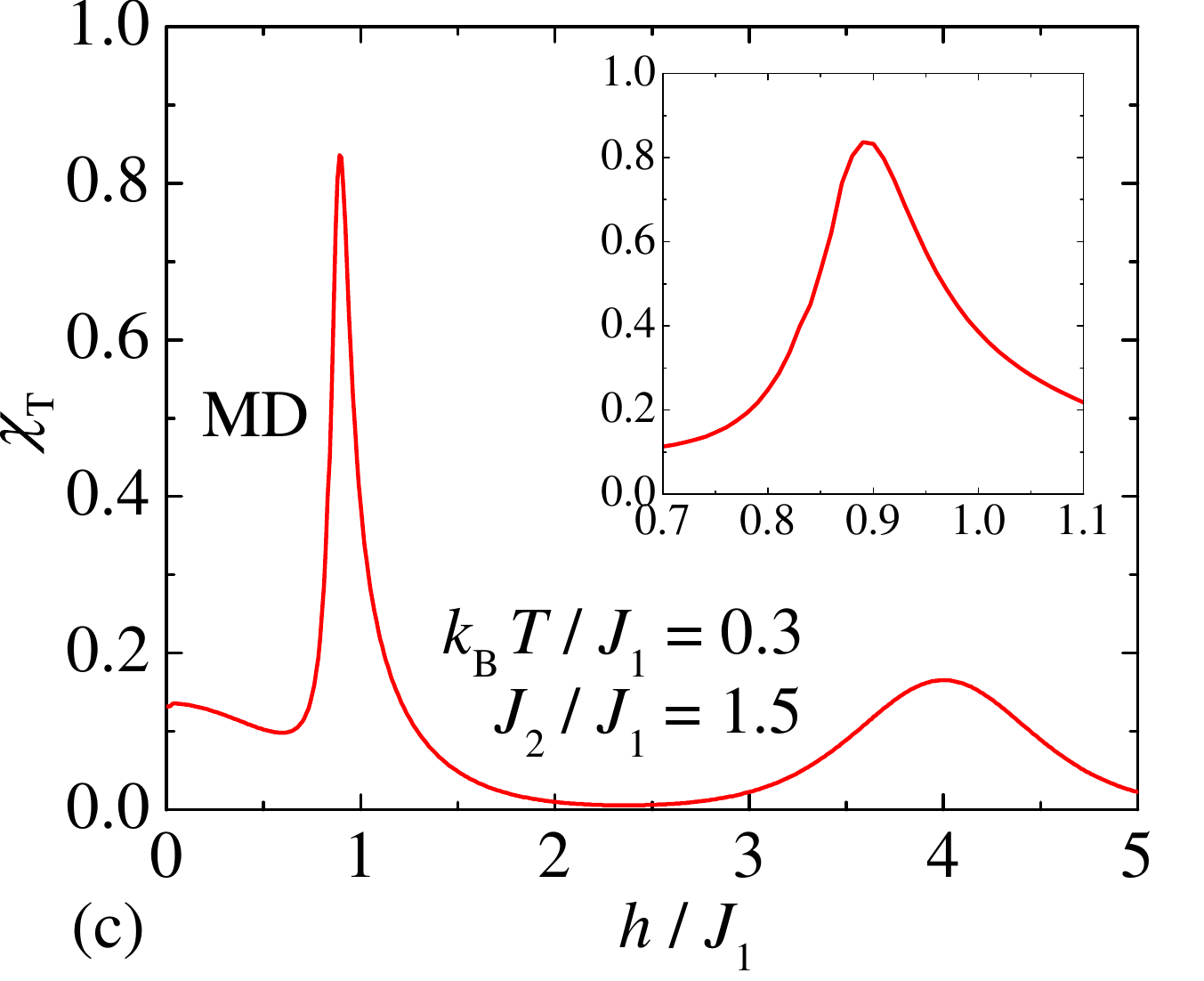}
\caption{A few typical magnetic-field dependencies of the isothermal magnetic susceptibility 
for the fixed value of the interaction parameter $J_2/J_1 =1.5$ and three selected values of temperature:
(a) $k_{\rm B}T/J_1 = 0.2$;
(b) $k_{\rm B}T/J_1 = 0.2403$;
(c) $k_{\rm B}T/J_1 = 0.3$.}
\label{msu}       
\end{figure}

Next, let us illustrate a few typical features of the magnetization and magnetic susceptibility when the magnetic field drives the spin-1/2 Ising-Heisenberg model on the diamond-like decorated lattice across the thermal phase transition. For this purpose, a few typical magnetic-field dependencies of the local and total magnetization are plotted in Fig.~\ref{mh} for three selected temperatures. The calculated values of the local magnetization of the Ising spins $m_{\rm I}$ (red lines) and the local magnetization of the Heisenberg spins 
$m_{\rm H}$ (blue lines) are indeed consistent with the presence of the MD and FRI phase for $h/J_{1} \lesssim 1$ and $h/J_{1} \gtrsim 1$, respectively. At the lowest temperature $k_{\rm B}T/J_1 = 0.2$ one detects a discontinuous thermal phase transition from the MD phase towards the FRI phase accompanied with an abrupt magnetization jump [see Fig.~\ref{mh}(a)]. The discontinuous magnetization jump gradually shrinks upon increasing temperature until it completely vanishes in vicinity of the critical temperature $k_{\rm B}T/J_1 \approx 0.2403$, where the magnetization discontinuity is replaced by an inflection point with a vertical (infinite) tangent [see Fig.~\ref{mh}(b)]. The inflection point finally acquires a finite slope at the even higher temperature $k_{\rm B}T/J_1 = 0.3$, where one finds a rather simple crossover from the MD phase towards the FRI phase. A similar crossover can be observed around the saturation field $h/J_1 = 4$, where the local and total magnetization undergo a substantial magnetic-field-driven change from the values typical for the FRI phase towards their fully saturated values.

All aforementioned features of the isothermal magnetization curves of the spin-1/2 Ising-Heisenberg diamond-decorated square lattice can be unambiguously corroborated by the magnetic-field-induced changes of the isothermal magnetic susceptibility presented in Fig.~\ref{msu}. The finite cusp of the magnetic susceptibility encountered at the lowest temperature $k_{\rm B}T/J_1 = 0.2$ verifies the existence of a discontinuous thermal phase transition between the MD and FRI phases [Fig.~\ref{msu}(a)]. On the other hand, there are strong indications that
in proximity of the critical temperature $k_{\rm B}T/J_1 \approx 0.2403$ of the continuous thermal phase transition
the magnetic susceptibility displays
a pronounced power-law divergence [Fig.~\ref{msu}(b)]. Last but not least, the magnetic susceptibility exhibits a rather sharp but rounded finite maximum at higher temperatures such as $k_{\rm B}T/J_1 = 0.3$ [see the inset in Fig.~\ref{msu}(c)]. It is also worth mentioning that the magnetic susceptibility displays for arbitrary temperature another round maximum located around $h/J_1 \approx 4$, which confirms a simple crossover from the FRI phase towards the fully saturated paramagnetic phase rather than a true phase transition.  

\begin{figure}[t!]
\includegraphics[width=\columnwidth]{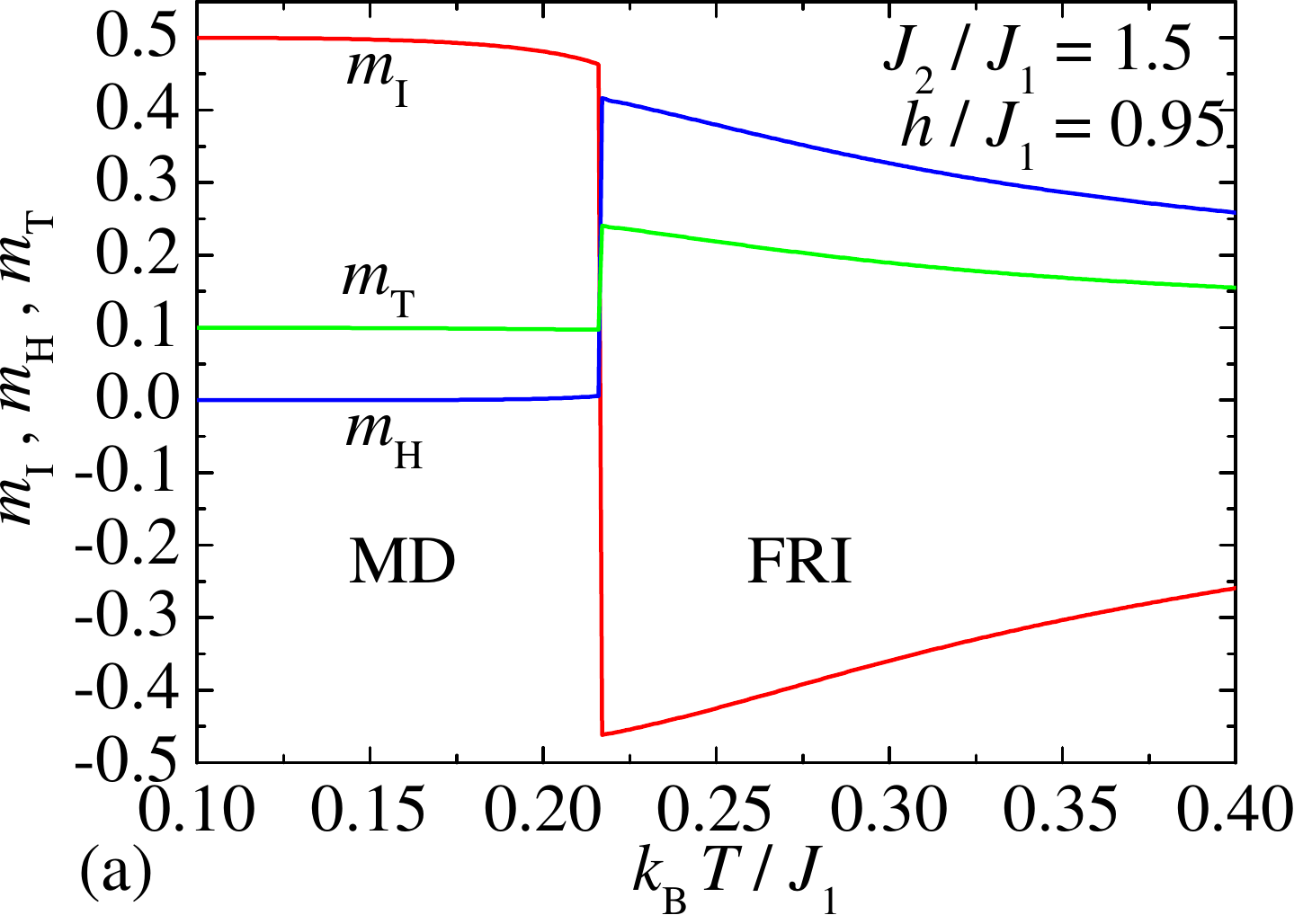}\\[1.5mm]
\includegraphics[width=\columnwidth]{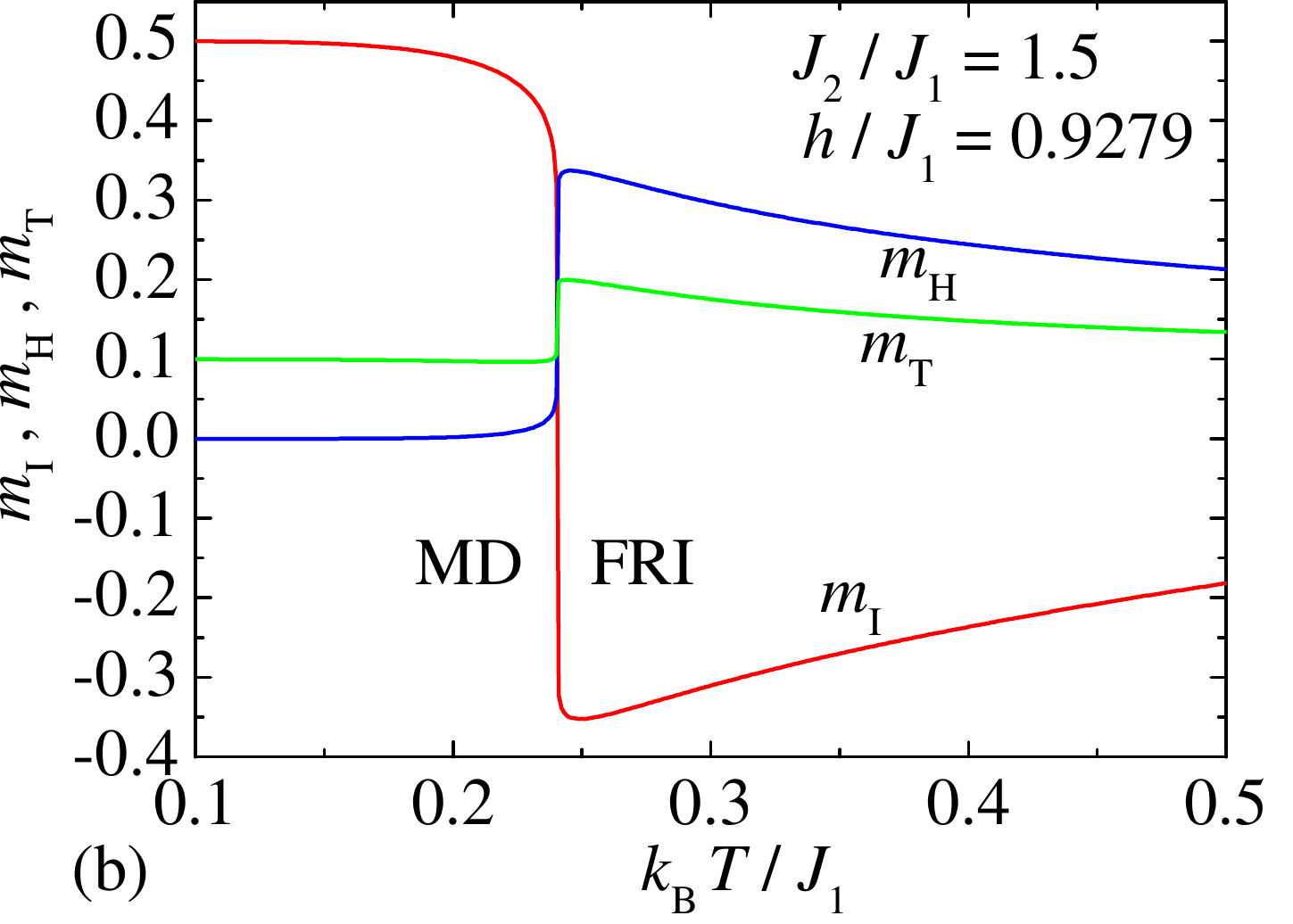}\\[1.5mm]
\includegraphics[width=\columnwidth]{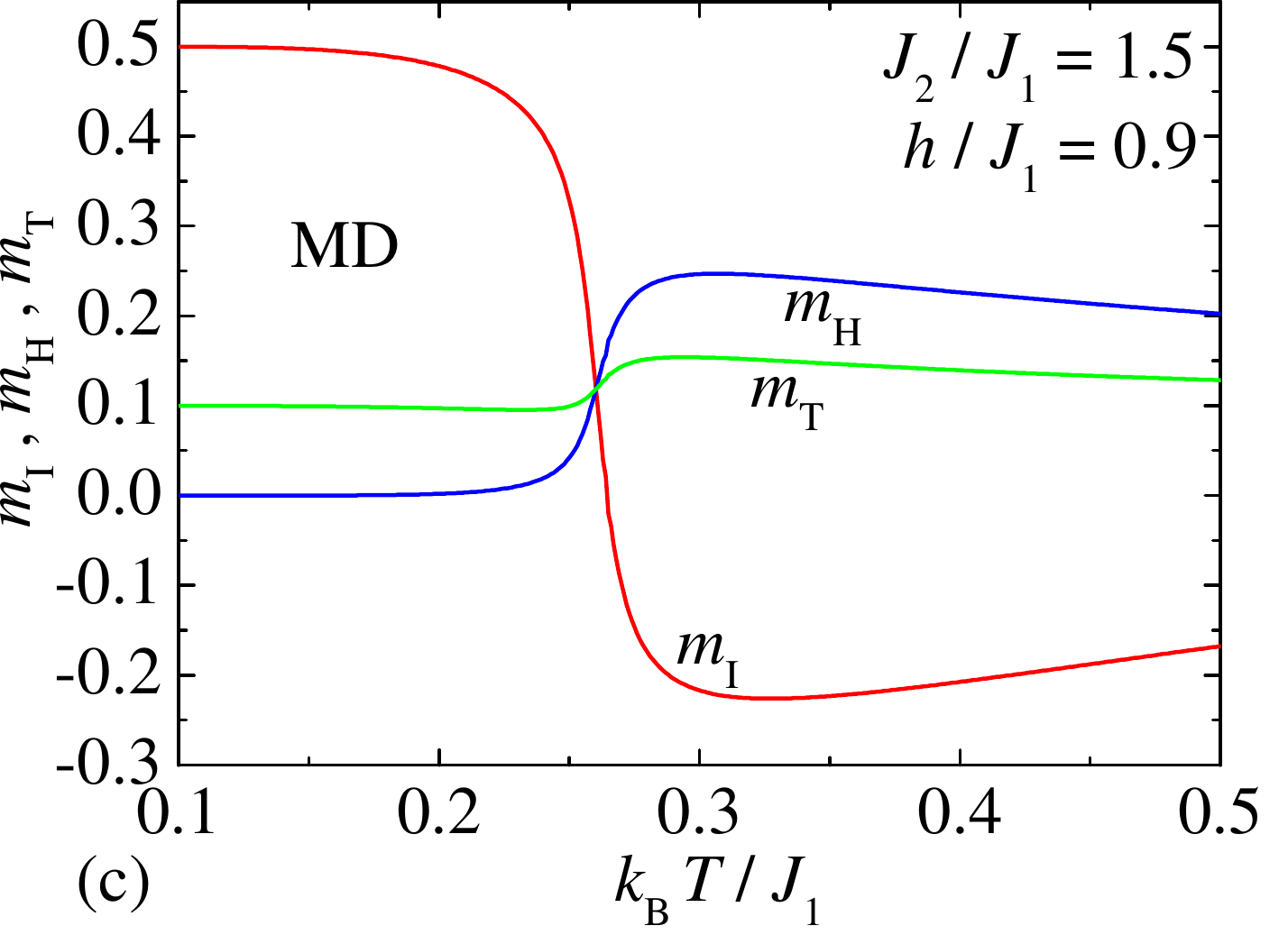}
\caption{A few typical temperature variations of the local and total magnetization 
for the fixed value of the interaction ratio $J_2/J_1 =1.5$ and three different magnetic fields:
(a) $h/J_1 = 0.95$;
(b) $h/J_1 = 0.9279$;
(c) $h/J_1 = 0.9$.}
\label{mt}       
\end{figure}

In the following we will examine in detail typical features of the magnetization and magnetic specific heat induced by \emph{temperature}, when driving the spin-1/2 Ising-Heisenberg model on the diamond-decorated square lattice across the thermal phase transitions.  To start with, Fig.~\ref{mt} depicts typical temperature variations of the local and total magnetization for three selected values of the magnetic field. The discontinuous nature of the local and total magnetization, observable at the selected value of the magnetic field $h/J_1 = 0.95$, corroborates the presence of a discontinuous thermal phase transition between the FRI and MD phases [Fig.~\ref{mt}(a)]. If the magnetic field is fixed sufficiently close to the critical value $h/J_1 \approx 0.9279$, the local and total magnetization show an inflection point with a vertical tangent serving in evidence of the continuous thermal phase transition between the FRI and MD phases [Fig.~\ref{mt}(b)]. At even lower magnetic field $h/J_1 = 0.9$ the local and total magnetization display a smooth continuous thermally-assisted change due to a crossover from the MD phase towards the FRI phase. 

\begin{figure}[t!]
\includegraphics[width=\columnwidth]{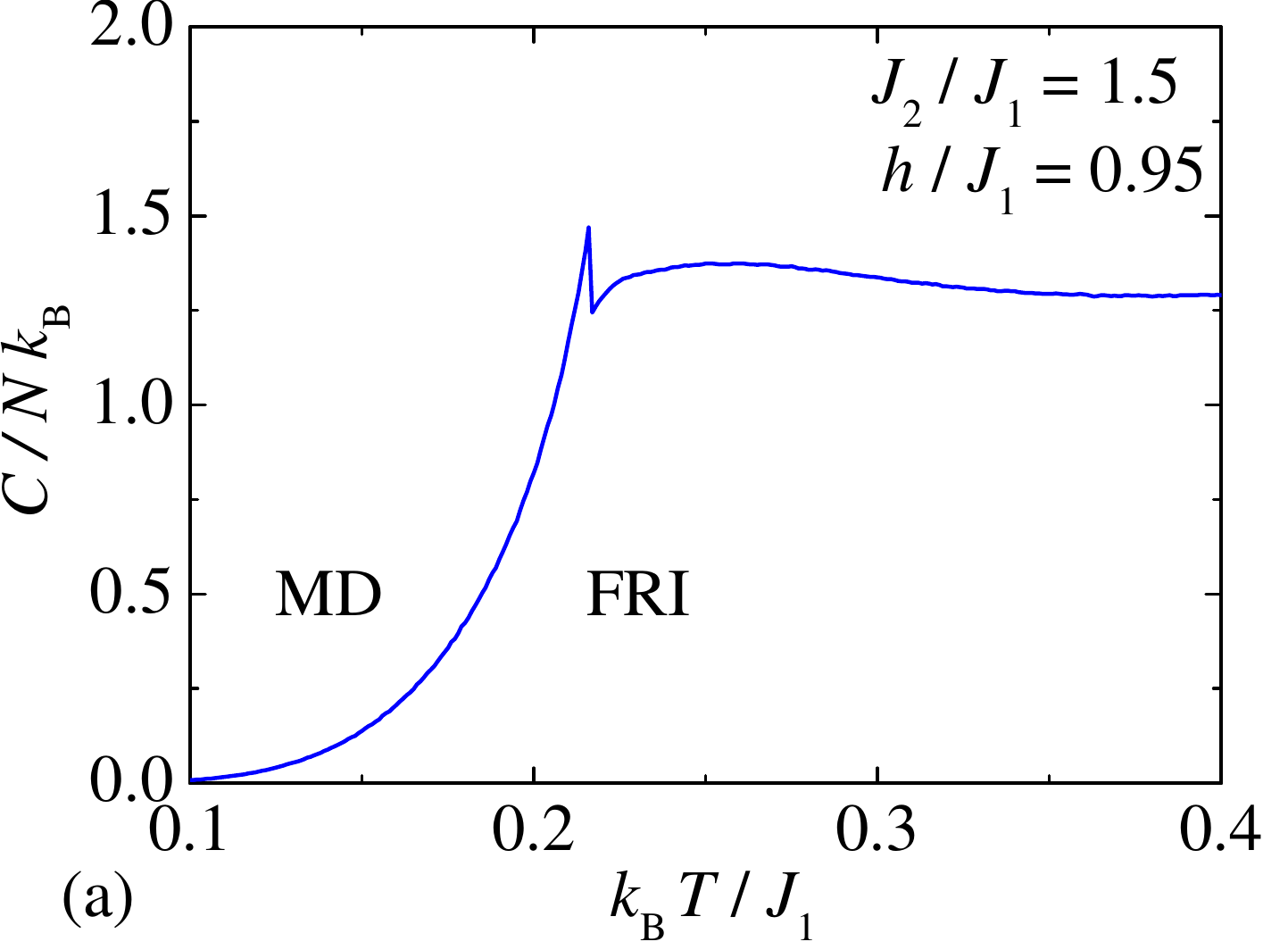}\\[1.5mm]
\includegraphics[width=\columnwidth]{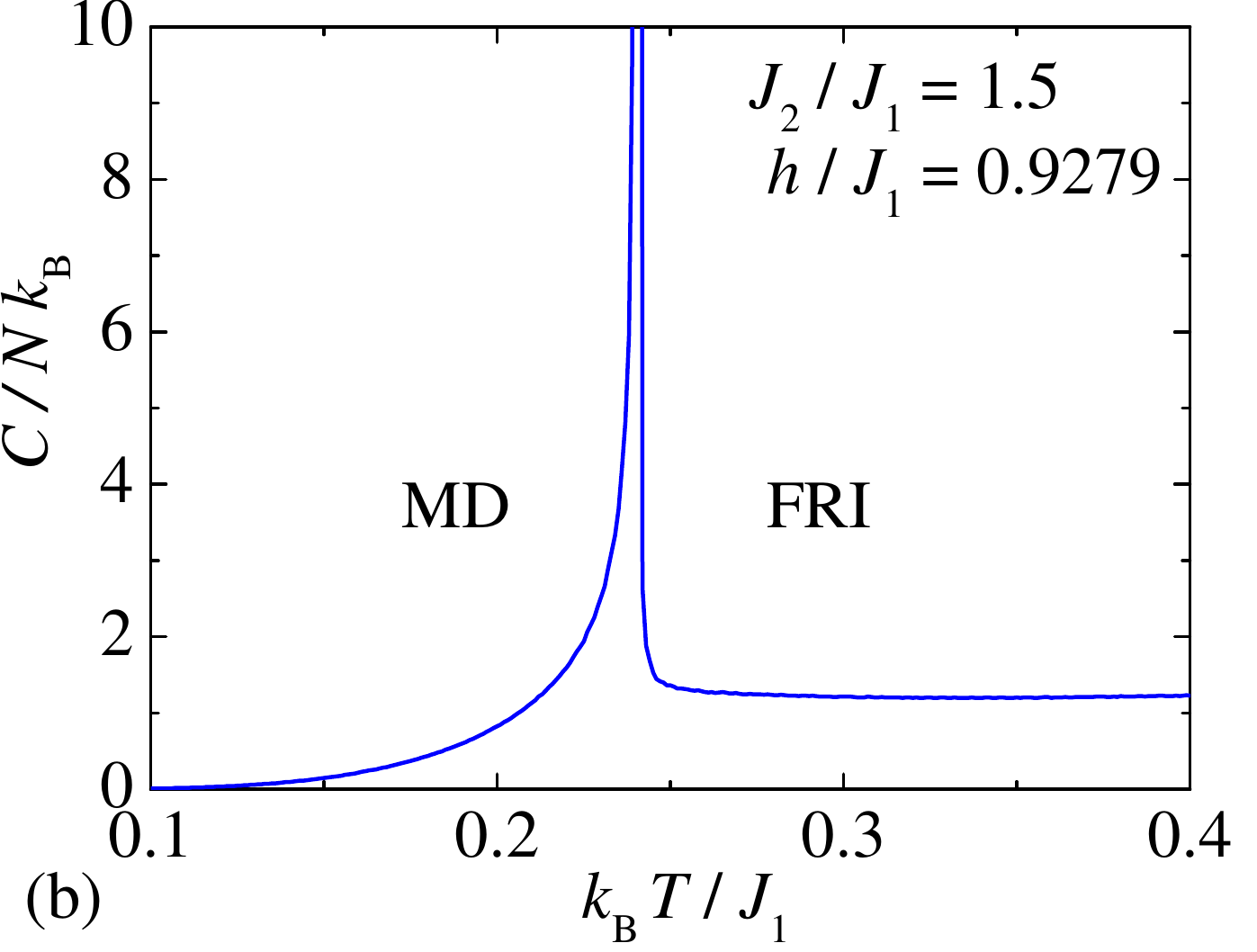}\\[1.5mm]
\includegraphics[width=\columnwidth]{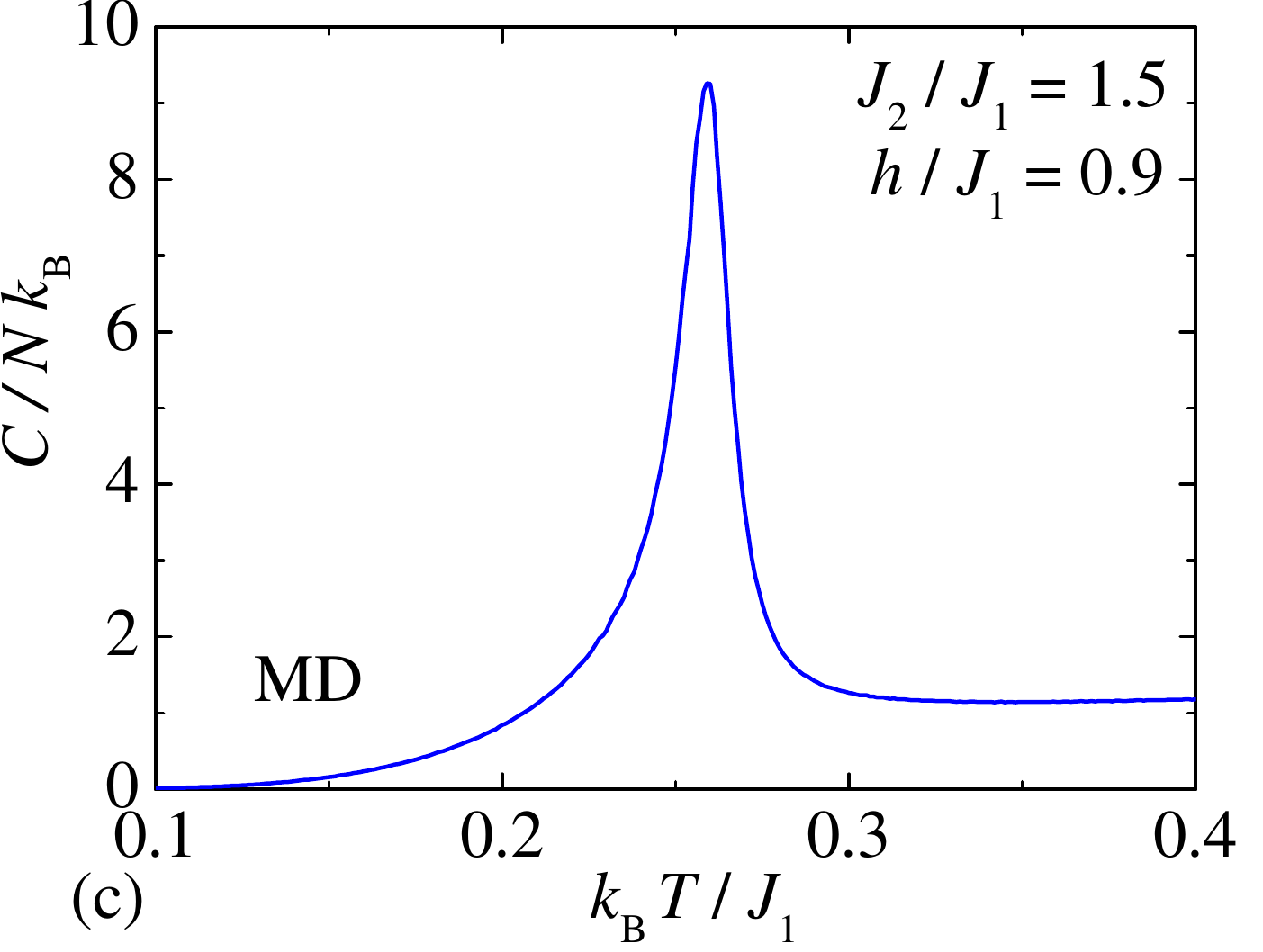}
\caption{A few typical temperature variations of the magnetic specific heat
for the fixed value of the interaction ratio $J_2/J_1 =1.5$ and three different magnetic fields:
(a) $h/J_1 = 0.95$;
(b) $h/J_1 = 0.9279$;
(c) $h/J_1 = 0.9$.}
\label{msh}       
\end{figure}
Let us conclude our discussion by a detailed analysis of the temperature dependencies of the magnetic specific heat of the spin-1/2 Ising-Heisenberg model on the diamond-decorated square lattice presented in Fig.~\ref{msh}.
The finite cusp of the magnetic specific heat seen in Fig.~\ref{msh}(a) around the temperature $k_{\rm B} T/J_1 \approx 0.216$ evidences the discontinuous thermal phase transition for the relevant choice of the magnetic field $h/J_1 = 0.95$. Contrary to this, the temperature dependence of the magnetic specific heat is quite reminiscent of a steep logarithmic divergence if the magnetic field $h/J_1 \approx 0.9279$ is tuned sufficiently close to the
continuous thermal phase transition [Fig.~\ref{msh}(b)]. Finally,
the magnetic specific heat may be completely free of any thermal phase transition as exemplified by the temperature dependence with a sizable but finite round maximum shown in Fig.~\ref{msh}(c) for the magnetic field $h/J_1 = 0.9$.

\section{CONCLUSION}

In the present article we have examined in detail discontinuous and continuous thermal phase transitions of the spin-1/2 Ising-Heisenberg model on the diamond-decorated square lattice between the classical FRI phase and the quantum MD phase in the presence of an external magnetic field. It has been demonstrated that the spin-1/2 Ising-Heisenberg diamond-decorated square lattice shows for a fixed value of the interaction ratio a line of discontinuous thermal phase transitions, which terminates at an Ising-type critical point inherent to a continuous thermal phase transition. It could be thus concluded that the phase boundary between the FRI and MD phases of the spin-1/2 Ising-Heisenberg model on the diamond-decorated square lattice is
reminiscent of the liquid-gas phase boundary of water. In addition, the results for discontinuous and continuous thermal phase transitions of the spin-1/2 Ising-Heisenberg diamond-decorated square lattice are exact as they can be directly descended from a rigorous mapping correspondence to an exactly solved spin-1/2 Ising model on a square lattice with a temperature-dependent nearest-neighbor interaction $J_{\rm eff} \neq 0$ and zero effective field $h_{\rm eff} = 0$ \cite{ons44}. Out of the parameter region corresponding to thermal phase transitions the spin-1/2 Ising-Heisenberg diamond-decorated square lattice can be rigorously mapped onto the spin-1/2 Ising model on a square lattice with a nonzero nearest-neighbor interaction $J_{\rm eff} \neq 0$ and effective field $h_{\rm eff} \neq 0$, which can be subsequently treated 
by classical Monte Carlo simulations.

By making use of extensive Monte Carlo simulations we have evidenced that the spin-1/2 Ising-Heisenberg model on the diamond-decorated square lattice displays a discontinuous magnetization jump when temperature or magnetic field drives the investigated spin system across a discontinuous thermal phase transition. The magnetic susceptibility and specific heat consequently exhibit at the discontinuous thermal phase transitions finite cusps related to a discontinuous change of both quantities. On the contrary, the magnetization of the spin-1/2 Ising-Heisenberg diamond-decorated square lattice varies continuously when temperature or magnetic-field changes drive the investigated spin system across a continuous thermal phase transition. Under this condition, the magnetic susceptibility and specific heat display a strong divergence at the continuous thermal phase transition.

The present accurate results are
reminiscent of the experimental findings, which recently reported unconventional thermal phase transitions of the two-dimensional quantum magnet SrCu$_2$(BO$_3$)$_2$ \cite{lar21}.
In the case of the diamond-decorated square lattice, the FRI-MD phase transition in the spin-1/2 Ising-Heisenberg
model is not only qualitatively but even quantitatively close to the Lieb-Mattis to MD phase transition
in the full spin-1/2 Heisenberg model \cite{cac23,str23}. The exact solution of the spin-1/2 Ising-Heisenberg
model thus permits us to unveil, for example, a reentrance in the finite-temperature transition line \cite{str23} that would be beyond the accuracy of the numerical methods used to solve the spin-1/2 Heisenberg model \cite{cac23}.

It is our hope that the present essentially exact results will stimulate further exploration in this exciting research field. 

\section{ACKNOWLEDGEMENT}
This work was funded by the Slovak Research and Development Agency and the French Ministry for Europe and Foreign Affairs, the French Ministry for Higher Education and Research under the \v{S}tef\'anik programme for Slovak-France bilateral projects under contract Nos.\ SK-FR-19-0013/45125RC and SK-FR-22-0011/49880PG. J.S.\ and K.K.\ acknowledge the partial financial support by a grant of the Slovak Research and Development Agency under the contract No.\ APVV-20-0150. We acknowledge support from the Deutsche Forschungsgemeinschaft (DFG) through Grant No.\ WE/3649/4-2 of the FOR 1807 and RTG 1995.


\begin{thebibliography}{99}
\leftskip=-5pt \vspace{-0.3truecm}

\bibitem{dom72} C.~Domb, M.~S.~Green: Phase Transitions and Critical Phenomena: Volume 1 (Academic Press, London - 1972); {\it ibid}. Volume 2 (Academic Press, London - 1972); 
{\it ibid}. Volume 3 (Academic Press, London - 1976); {\it ibid}. Volume 4 (Academic Press, London - 1977); {\it ibid}. Volume 5 (Academic Press, London - 1983).
\bibitem{lar21} J.~Larrea~Jim\'enez, S.~P.~G.~Crone, E.~Fogh, M.~E.~Zayed, R.~Lortz, E.~Pomjakushina, K.~Conder, A.~M.~L\"auchli,
                L.~Weber, S.~Wessel, A.~Honecker, B.~Normand, Ch.~R\"uegg, P.~Corboz, H.~M.~Ronnow, F.~Mila, Nature \textbf{592}, 370 (2021).
\bibitem{sta18} J.~Stapmanns, P.~Corboz, F.~Mila, A.~Honecker, B.~Normand, S.~Wessel, Phys. Rev. Lett. \textbf{121}, 127201 (2018).
\bibitem{web22} L.~Weber, A.~Y.~D.~Fache, F.~Mila, S.~Wessel, Phys. Rev. B \textbf{106}, 235128 (2022).
\bibitem{fan23} Y.~Fan, N.~Xi, C.~Liu, B.~Normand, R.~Yu, arXiv: 2306.16288.
\bibitem{hon22} L.~Weber, A.~Honecker, B.~Normand, P.~Corboz, F.~Mila, S.~Wessel, SciPost Phys. \textbf{12}, 054 (2022).
\bibitem{cac23} N.~Caci, K.~Karl'ov\'a, T.~Verkholyak, J.~Stre\v{c}ka, S.~Wessel, A.~Honecker, Phys. Rev. B {\bf 107}, 115143 (2023).	
\bibitem{gub16} J.~Gubernatis, N.~Kawashima, P.~Werner: Quantum Monte Carlo Methods: Algorithms for Lattice Models (Cambridge University Press, Cambridge - 2016). 		
\bibitem{str23} J.~Stre\v{c}ka, K.~Karl'ov\'a, T.~Verkholyak, N.~Caci, S.~Wessel, A.~Honecker, Phys. Rev. B {\bf 107}, 134402 (2023).

\bibitem{bax82} R.~J.~Baxter: Exactly Solved Models in Statistical Mechanics (Academic, London, 1982).
\bibitem{fis60} M.~E.~Fisher, Proc. R. Soc. London A \textbf{254}, 66 (1960); {\it ibid}. \textbf{256}, 502 (1960).
\bibitem{hat68} M.~Hattori, H.~Nakano, Prog. Theor. Phys. \textbf{40}, 958 (1968).
\bibitem{mas73} H.~Mashiyama, S.~Nara, Phys. Rev. B \textbf{7}, 3119 (1973).
\bibitem{luw05} W.~T.~Lu, F.~Y.~Wu, Phys. Rev. E \textbf{71}, 046120 (2005).
\bibitem{can06} L.~\v{C}anov\'a, M.~Ja\v{s}\v{c}ur, Condens. Matter Phys. \textbf{9}, 47 (2006).
\bibitem{gal16} L.~G\'alisov\'a, J. Phys.: Condens. Matter \textbf{28}, 476005 (2016).

\bibitem{fis59} M.~E.~Fisher, Phys. Rev. \textbf{113}, 969 (1959).
\bibitem{roj09} O.~Rojas, J.~S.~Valverde, S.~M.~de~Souza, Physica A, \textbf{388}, 1419 (2009). 
\bibitem{str10} J.~Stre\v{c}ka, Phys. Lett. A \textbf{374}, 3718 (2010). 
\bibitem{ons44} L.~Onsager, Phys. Rev. \textbf{65}, 117 (1944). 

\end{thebibliography}
\end{document}